\documentclass[aps,prb,twocolumn,groupedaddress]{revtex4-1}

\usepackage{graphicx}
\usepackage{verbatim}
\begin{document}

\title{NMR Characterization of Sulphur Substitution Effects in the K$_{x}$Fe$_{2-y}$Se$_{2-z}$S$_{z}$ high-$T_{c}$ Superconductor}

\author{D. A. Torchetti$^{1}$}
\author{T. Imai$^{1,2}$}
\author{H. C. Lei $^{3}$}
\author {C. Petrovic $^{3}$}
\affiliation{$^{1}$ Department of Physics and Astronomy, McMaster University, Hamilton, Ontario, L8S 4M1, Canada}
\affiliation{$^{2}$ Canadian Institute for Advanced Research, Toronto, Ontario, M5G 1Z8, Canada}
\affiliation{$^{3}$ Condensed Matter Physics and Materials Science Department, Brookhaven National Laboratory, Upton, New York, 11973, USA}

\date{\today}

\begin{abstract}
We present a $^{77}$Se NMR study of the effect of S substitution in the high $T_{c}$ superconductor, K$_{x}$Fe$_{2-y}$Se$_{2-z}$S$_{z}$, in a temperature range up to 250 K. We examine two S concentrations, with z = 0.8 ($T_{c} \sim$ 26 K) and z = 1.6 (non-superconducting).  The samples containing sulphur exhibit broader NMR lineshapes than the K$_{x}$Fe$_{2}$Se$_{2}$ sample due to local disorder in the Se environment.  Our Knight shift $^{77}K$ data indicate that in all samples uniform spin susceptibility decreases with temperature, and that the magnitude of the Knight shift itself decreases with increased S concentration.  In addition, S substitution progressively suppresses low frequency spin fluctuations.  None of the samples exhibit an enhancement of low frequency antiferromagnetic spin fluctuations (AFSF) near $T_{c}$ in 1/T$_{1}$T, as seen in FeSe.  

\end{abstract}

\maketitle


\section{Introduction}
With the discovery of high $T_{c}$ superconductivity in the iron-arsenides in 2008,~\cite{Hosono} researchers found renewed interest and hope in understanding the mechanism of high $T_{c}$ superconductivity.~\cite{Norman,Day,Paglione,Johnston}  Although much has been learned, the superconducting mechanism is still poorly understood.  Earlier reports suggest that optimal doping for superconductors places the system in the vicinity of magnetic instability as evidenced by the largest enhancement of antiferromagnetic spin fluctuations (AFSF) near $T_{c}$.~\cite{Nakai,Ning1,Ning2,Ning3}  Fermi surface nesting between hole bands at the Brillouin zone center and electron bands near the zone edge may be associated with the AFSF, which in turn may cause Cooper pairing.  In such a scenario, overdoping electrons suppresses the nesting effects by filling up the hole bands, and thereby suppressing AFSF and $T_{c}$.~\cite{Ning2,Ning3}  In the simple iron-selenide superconductor, FeSe ($T_{c} \sim$ 9 K~\cite{Hsu,McQueen,Mizuguchi4}), application of pressure alone can also push $T_{c}$ as high as 37 K.~\cite{Medvedev,Margadonna,Mizuguchi}  Furthermore, AFSF were shown to enhance in direct correlation with the enhancement of pressure and $T_{c}$,~\cite{Imai} reinforcing the view that AFSF and the superconducting mechanism may be linked.  

Recently, Guo \textit{et al.} reported the discovery of superconductivity in K$_{x}$Fe$_{2-y}$Se$_{2}$.~\cite{Guo}  Through intercalation of K into FeSe, $T_{c}$ is again pushed much higher ($\sim$ 33 K).  Recent ARPES measurements, however, revealed that unlike in iron-arsenide high $T_{c}$ systems, \textit{all} the hole bands near the center of the first Brillouin zone may be filled by electrons donated by K$^{+}$ ions.~\cite{Qian}  This means that Fermi surface nesting effects would not induce AFSF, yet $T_{c}$ is still very high.  In fact, Earlier NMR results revealed no enhancement of the AFSF toward $T_{c}$ in this system.~\cite{Yu,Kotegawa,Me} 

In the present study, we explore the effect of substituting sulphur (S) into the Se sites of K$_{x}$Fe$_{2-y}$Se$_{2-z}$S$_{z}$.~\cite{Petrovic4,Petrovic1}  As shown in the phase diagram in Fig. 1, S substitution progressively suppresses $T_{c}$, enabling us to conduct a systematic microscopic NMR investigation of the evolution of electronic properties in going from a high-$T_{c}$ superconductor to a non-superconductor.  We examine two concentrations, one with z = 0.8 (40\% substitution), and one with z = 1.6 (80\% substitution), and compare the new NMR results with our report for K$_{x}$Fe$_{2-y}$Se$_{2}$ (z = 0).~\cite{Me}  Sulphur substitution tends to generate a chemical pressure in the system, because S and Se have the same valence but S has an ionic radius less than half of that of Se.  The lattice constant along the crystal $c$-axis indeed decreases from $\sim$14.0 to 13.9 \AA\, and then to 13.7 \AA\  as S concentration goes from z = 0 to 0.8 to 1.6, respectively; the lattice is compressed within the $ab$ plane as well, so overall the Fe-Se bond length is shortened.~\cite{Petrovic1}   The systematic suppression of $T_{c}$ with S substitution mimics the situation in FeSe, where over-pressurizing results in suppression of $T_{c}$.~\cite{Medvedev}  The evolution of electronic properties in overpressured FeSe has been poorly explored because extremely high pressure ($\geq$ 9 GPa) is required.~\cite{Medvedev}  Thus a microscopic investigation of S-substituted K$_{x}$Fe$_{2-y}$Se$_{2}$ may provide us with a new path to investigate the possible correlations between $T_{c}$, structure, and AFSF when physical or chemical pressure suppresses $T_{c}$.

\begin{figure}[h]
\includegraphics[width=3in, height=3.4in, angle=270]{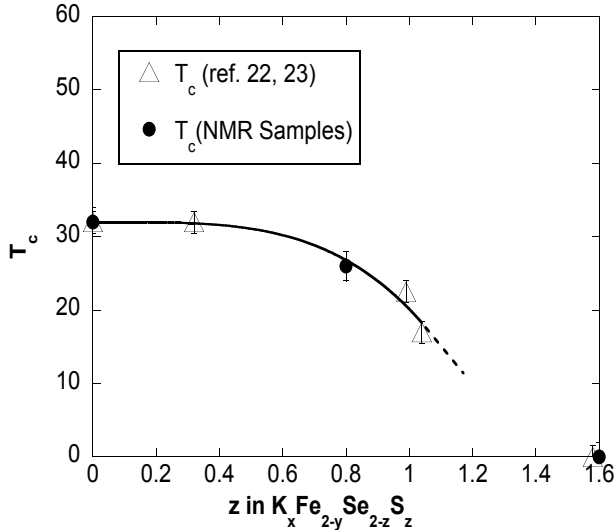}
\caption{(Color Online) Superconducting transition temperature, $T_{c}$, as a function of sulphur concentration, z.  Filled circles represent current samples used for NMR, measured \textit{in situ} in zero magnetic field, while open triangles represent data at the source of crystal growth.~\cite{Petrovic4}  Solid and dashed lines are a `guide-to-eye'.}
\end{figure}

In what follows we will present new $^{77}$Se NMR data for K$_{x}$Fe$_{2-y}$Se$_{2-z}$S$_{z}$, and show that spin excitations are suppressed with S substitution.  In section II we will briefly present our experimental procedures.  In section III we will summarize our experimental results, paying particular attention to NMR Knight shift and nuclear spin-lattice relaxation rate 1/T$_{1}$ data, and discuss some implications.  In section IV we summarize our findings and conclude.  


\section{Experimental}
Single crystals of K$_{x}$Fe$_{2-y}$Se$_{2-z}$S$_{z}$ were grown using the self-flux method.  Powder X-ray diffraction (XRD) measurements were obtained at 300 K using 0.3184 \AA\  wave-length radiation on ground samples.
The XRD data were then refined with the General Structure Analysis System.~\cite{GSAS,Toby}  Average stoichiometries were determined using energy-dispersive X-ray spectroscopy (EDX) in a JEOL JSM-6500 scanning electron microscope.  The XRD data show superlattice reflections which incorporate an Fe ordered vacancy site.  Potassium content as well as Fe defect content increase slightly with S substitution, as $x \sim$ 0.64, 0.7, and 0.8 in K$_{x}$Fe$_{2-y}$Se$_{2}$, K$_{x}$Fe$_{2-y}$Se$_{1.2}$S$_{0.8}$, and K$_{x}$Fe$_{2-y}$Se$_{0.4}$S$_{1.6}$ respectively.  Full detail of the growth method, structural refinement, phase diagram, and other properties may be found elsewhere.~\cite{Petrovic4,Petrovic1,Petrovic3,Petrovic2}  We determined the $T_{c}$ of each of our NMR samples \textit{in situ} by measuring the frequency shift of our NMR tank circuit.~\cite{Me}  The $T_{c}$ of our NMR samples is summarized in Fig. 1, together with other compositions. 

We carried out NMR measurements in an applied magnetic field of B $= 8.33$ Tesla applied along the crystal $c$-axis.  By taking the Fast Fourier Transformation (FFT) of the envelope of the spin-echo we were able to obtain our NMR lineshapes.  To measure the nuclear spin-lattice relaxation rate, 1/T$_{1}$, we applied an inversion $\pi$ pulse or saturation comb pulses prior to our spin-echo sequence.

  
\section{Results and Discussion}
\subsection{Uniform Spin Susceptibility}

In Fig. 2 we compare $^{77}$Se NMR lineshapes for different sulphur concentrations.  Once scaled according to Selenium content in each crystal, integrated intensities for each system are equal within experimental uncertainties.  The narrowest lineshapes are observed for the $z=0$ sample; both S substituted samples exhibit line broadening, indicative of disorder effects in local electronic and structural properties at the Se sites. 

\begin{figure}[h]
\includegraphics[width=3in, height=3.4in, angle=270]{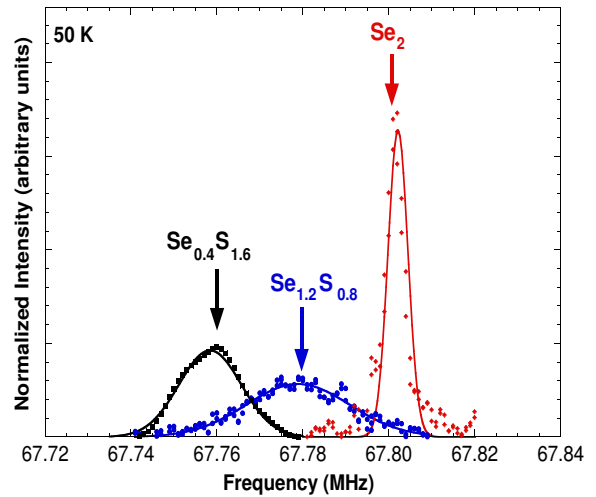}
\caption{(Color Online) Representative $^{77}$Se NMR FFT lineshapes of K$_{x}$Fe$_{2-y}$Se$_{2-z}$S$_{z}$ for z = 0.8 and z = 1.6 at 50 K, observed in B = 8.33 T.  We also show K$_{x}$Fe$_{2-y}$Se$_{2}$ for comparison.~\cite{Me}  The lineshapes have been normalized for an equal integrated intensity and fitted with Gaussian curves with Full Widths at Half Maximum (FWHM) $\sim$ 4.5 KHz, 24 KHz, and 14 KHz for z = 0, 0.8, and 1.6, respectively. Note the systematic lowering of the NMR peak frequency with increased S content.}
\end{figure}

In addition to the change in shape of the FFT, the NMR peak frequency is systematically lowered with increasing $z$.   Peaks for the z = 0.8 and z = 1.6 samples are found at 67.78 MHz and 67.76 MHz respectively.  With an expected Larmor frequency of $f_{0}$ = $\gamma_{n}$ B = 67.622 MHz, where $\gamma_{n}$/2$\pi$ = 8.118 MHz/T is the \textit{gyromagnetic ratio} for $^{77}$Se, this means that the \textit{Knight Shift} is lowered with increasing S concentration as well.  

The $^{77}$Se NMR Knight shift, $^{77}K$, may be written as
\begin{equation}
^{77}K = \frac{A_{hf}}{g\mu_{B}}\chi_{spin} + K_{chem}.
\end{equation}
$A_{hf}$ is the hyperfine coupling constant between the $^{77}$Se nuclear spins which we are observing and electron spins in their vicinity.  $\chi_{spin}$ is the average spin susceptibility of these electrons.  The Knight shift is therefore an excellent probe for the uniform \textbf{q = 0} wave vector mode of local spin susceptibility in bulk materials, free from the influence of impurity phases.  $K_{chem}$ is a temperature independent chemical shift which arises from the orbital motion of electrons. From the measurements below $T_{c}$ in K$_{x}$Fe$_{2}$Se$_{2}$ we estimate the lower bound of $K_{chem}$ to be $\sim$ 0.11 \% without diamagnetic corrections.~\cite{Me}  $K_{chem}$ may be slightly dependent on S-substitution effects.
 
In Fig. 3 we summarize the temperature and S-content dependence of $^{77}K$.  We find some qualitative similarities in the behaviour of all three compositions, but the magnitude of the spin contribution to the shift is reduced significantly in the S substituted samples near room temperature.  In principle, the overall suppression of the spin contribution $^{77}K_{spin}$ = $\frac{A_{hf}}{g\mu_{B}}\chi_{spin}$ may be caused by a reduction in the hyperfine coupling constant, $A_{hf}$, induced by structural changes.  However, the Fe-Se bond length shortens as one substitutes S,~\cite{Petrovic1} which would cause greater overlap of wavefunctions between Fe and Se(S) layers.  Naively, $A_{hf}$ would therefore become larger.  Since we see an overall \textit{decrease} in Knight shift, this implies $\chi_{spin}$ would have to drop more substantially if $A_{hf}$ indeed increases.  That is, our results in Fig. 3 imply that S substitution suppresses $\chi_{spin}$.

Turning our attention to the temperature dependence of $^{77}K$, all samples exhibit a monotonic decrease in Knight shift with temperature, which is consistent with behaviour previously reported in other iron-based superconductors such as  LaFeAsO$_{1-x}$F$_{x}$,~\cite{Ahilan,Nakai} Ba(Fe$_{1-x}$Co$_{x}$)$_{2}$As$_{2}$,~\cite{Ning1,Ning2,Ning3,Wang}, FeSe,~\cite{Imai}, and Ba$_{x}$K$_{1-x}$Fe$_{2}$As$_{2}$.~\cite{Matano}  In Ba(Fe$_{1-x}$Co$_{x}$)$_{2}$As$_{2}$ the temperature dependence of Knight shift remains qualitatively similar regardless of doping level.~\cite{Ning3,Ning2} In the present case, however, the temperature dependence becomes far less pronounced with S substitution.  The increase in Knight shift at higher temperature for the undoped (z = 0) sample may be caused by the presence of excited states of the spins, which may be reached at higher temperatures with  sufficient thermal energy.  In such a pseudo-gap scenario,~\cite{Ahilan,Nakai} our new observation implies that in the S-substituted systems either these excited states are diminished or the energy threshold required to excite spins is too great (and beyond our temperature range).  

\begin{figure}[h]
\includegraphics[width=3in, height=3.4in, angle=270]{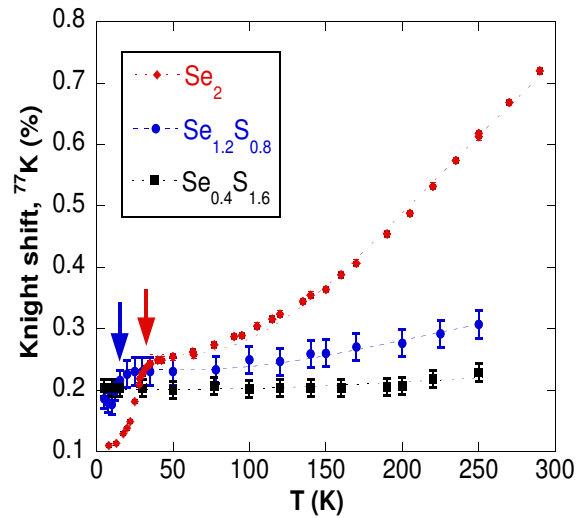}
\caption{(Color Online) Temperature dependence of the $^{77}$Se NMR Knight shift, $^{77}K$, along the $c$-axis.  K$_{x}$Fe$_{2-y}$Se$_{2}$ is presented for comparison.  Arrows mark $T_{c}$ = 25 K (z = 0) and 13 K (z = 0.8) in B = 8.33 T for the two superconducting samples as determined \textit{in situ} by the change in tuning frequency of the NMR tank circuit.  The dashed curve is a phenomenological fit with a pseudo-gap, $^{77}K \sim \alpha + \beta\exp{(-\Delta /k_{B}T)}$, with $\Delta/k_{B}$ = 435 K for all samples.  Overall, $^{77}K$ and therefore the \textbf{q = 0} component of $\chi_{spin}$ are suppressed with increased S doping.}
\end{figure} 

The dashed line through each of the three samples represents a purely phenomenological fit to the normal state data, with $^{77}K \sim \alpha + \beta\exp{(-\Delta /k_{B}T)}$, where $\Delta$ represents a pseudo-gap, $\alpha$ represents the $^{77}K$ value near $T_{c}$, and $\beta$ is a constant.  We had hoped to find a systematic change in $\Delta$ with S substitution but the uncertainty in its magnitude is too large because it is extremely sensitive to the choice of $\alpha$.   In fact, \textit{all three} samples can be fit with the same $\Delta/k_{B}$ = 435 K as previously found for undoped K$_{x}$Fe$_{2-y}$Se$_{2}$.~\cite{Me}  We emphasize that $\Delta$ can be varied by as much as a factor of 2 while still maintaining a good fit to the data, and so it does not provide sufficient resolution to discern the difference between the pseudo-gaps.  We have previously encountered this situation in Ba(Fe$_{1-x}$Co$_{x}$)$_{2}$As$_{2}$.~\cite{Ning2}

Within a conventional Fermi liquid picture, the Knight shift would be dominated by Pauli spin susceptibility, $\chi_{spin} = \mu_{B}^{2}N(E_{F})$.  A decrease in $^{77}K$ therefore might suggest that the density of states at the Fermi energy, $N(E_{F})$, decreases as the concentration of S increases and temperature decreases.  To check such a scenario beyond the \textbf{q = 0} mode of spin susceptibility, we must examine the nuclear spin-lattice relaxation rate, 1/T$_{1}$.


\subsection{Low Frequency Spin Fluctuations}
Fig. 4 shows representative signal recovery after saturation recorded for T$_{1}$ measurements for all three samples. Single exponential fits are satisfactory for all compositions, as expected for $^{77}$Se with nuclear spin I = 1/2.  We can clearly see the general trend that substitution of S slows the spin-lattice relaxation.  In Fig. 5 we present 1/T$_{1}$T (i.e. 1/T$_{1}$ divided by T) as a function of temperature.  1/T$_{1}$T measures the wave-vector \textbf{q} integral of the imaginary part of the dynamical spin susceptibility, $\chi^{"}$(\textbf{q},f$_{NMR}$), in the first Brillouin zone;
\begin{equation}
1/T_{1}T \propto \Sigma_{q}|A_{hf}(\textbf{q})|^{2}\frac{\chi^{"}(\textbf{q},f_{NMR})}{f_{NMR}},
\end{equation}
where $A_{hf}(\textbf{q})$ is the wave vector dependent hyperfine form factor, and $f_{NMR} \sim$ 67.78 MHz is the resonance frequency.   In the optimally doped regime, many iron-arsenide systems show an enhancement of low frequency AFSF as reflected by 1/T$_{1}$T as one approaches $T_{c}$ with decreasing temperature.~\cite{Nakai,Ning1,Ning2,Ning3,Matano}  The FeSe system exhibits an analogous enhancement of AFSF as well, and a greater enhancement in AFSF reflected by 1/T$_{1}$T correlates with higher T$_{c}$.~\cite{Imai}  For Ba(Fe$_{1-x}$Co$_{x}$)$_{2}$As$_{2}$, once the system enters the overdoped regime the enhancement of AFSF and $T_{c}$ both begin to decrease.~\cite{Ning2}

\begin{figure}[h]
\includegraphics[width=3in, height=3.4in, angle=270]{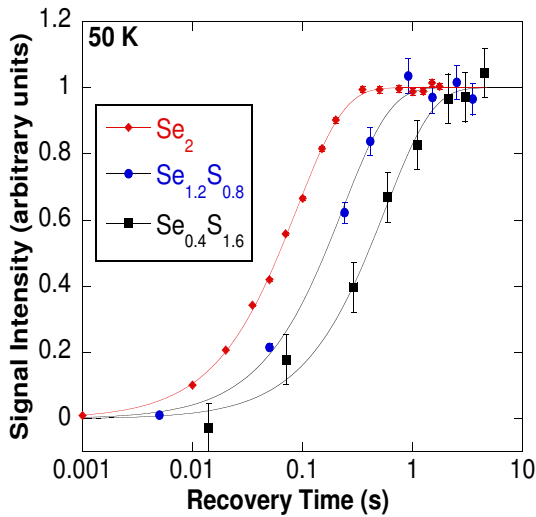}
\caption{(Color Online) Representative T$_{1}$ recovery curves measured at the center of the FFT lineshapes in Fig. 1 at 50 K.  We normalize signal intensity for clarity.  Solid lines represent single exponential fits through each data set, the argument of which gives our T$_{1}$ value.  The strong trend toward slower relaxation rate with increased S substitution is evident.}
\end{figure}

\begin{figure}[h]
\includegraphics[width=3in, height=3.4in, angle=270]{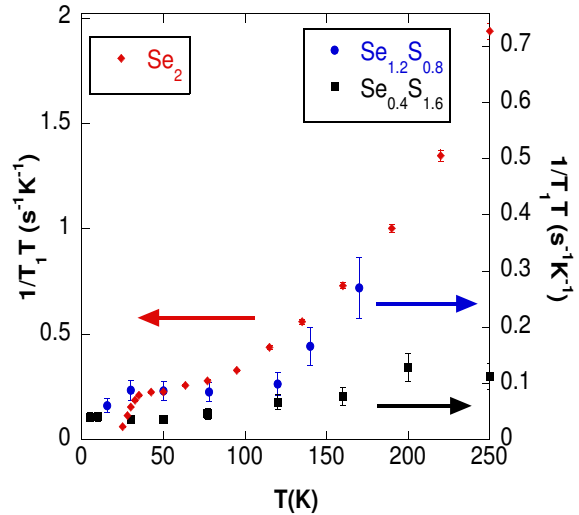} 
\caption{(Color Online) 1/T$_{1}$T as a function of temperature for z = 0, 0.8, and 1.6, with magnetic field B = 8.33 T applied along the crystal $c$-axis.  Relaxation rate gets progressively slower with S substitution; temperature dependence weakens as well.  Neither of the two superconducting samples exhibit an enhancement toward $T_{c}$, and S substituted samples show nearly flat behaviour except for near room temperature.}
\end{figure} 

In K$_{x}$Fe$_{2-y}$Se$_{2}$, 1/T$_{1}$T decreases sharply with temperature below 300 K, then levels off toward $T_{c}$.~\cite{Yu,Kotegawa,Me}  The lack of enhancement in 1/T$_{1}$T near $T_{c}$ is analogous to the case of overdoped Ba(Fe$_{1-x}$Co$_{x}$)$_{2}$As$_{2}$ and suggests that K$_{x}$Fe$_{2-y}$Se$_{2}$ may already be in the overdoped regime.  In K$_{x}$Fe$_{2-y}$Se$_{1.2}$S$_{0.8}$ we observed an analogous trend with overall suppression of 1/T$_{1}$T, and hence spin excitations.  Thus suppression of T$_{c}$ with S substitution in Fig. 1 is accompanied by suppresion of spin excitations.  For non-superconducting K$_{x}$Fe$_{2-y}$Se$_{0.4}$S$_{1.6}$, 1/T$_{1}$T is nearly temperature independent, similar to overdoped, non-superconducting, metallic Ba(Fe$_{0.7}$Co$_{0.3}$)$_{2}$As$_{2}$.~\cite{Ning2}  We caution that our finding does not necessarily imply there are no antiferromagnetic correlations in these systems.  Our results in Fig. 5 do not rule out the possible enhancement of AFSF at energy scales much higher than $\hbar f_{NMR} \sim  \mu$eV.  In addition, 1/T$_{1}$T measures the \textit{summation} of the wave-vector \textbf{q}, $\Sigma_{q}$, meaning that small AFSF may be present for some \textbf{q}-modes, and only that the net sum shows no enhancement.

\begin{figure}[h]
\includegraphics[width=3in, height=3.4in, angle=270]{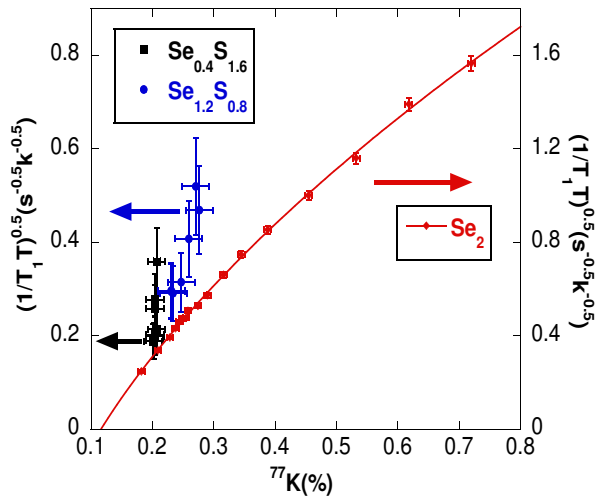}
\caption{(Colour Online) 1/$\sqrt{T_{1}T}$ vs $^{77}K(\%)$.  A constant slope would indicate the presence of the Korringa relation, an indicator for Fermi liquid theory.  The extrapolated horizontal intercept corresponds to $K_{chem}$.  The solid line through the undoped sample is a free parameter fit for 1/T$_{1}$T = ($^{77}K - K_{chem}$)$^{n}$, where $n \sim$ 1.6 instead of the Fermi liquid value of $n \sim$ 2, with the horizontal intercept $K_{chem} \sim$ 0.11\%.} 
\end{figure}          

In Fig. 6 we plot 1/$\sqrt{T_{1}T}$ vs $^{77}K(\%)$.  In a canonical Fermi liquid, 1/$T_{1}T \propto N(E_{F})^{2}$ as a result of Fermi's golden rule.  The Korringa relation, 
\begin{equation}
1/\sqrt{T_{1}T} = C\: (^{77}K_{spin}) + K_{chem},
\end{equation} 
where C is a constant and $^{77}K_{spin} \equiv \frac{A_{hf}}{g\mu_{B}}\chi_{spin}$ represents the spin contribution to Knight shift, is thus an excellent test for the applicability of Fermi liquid theory in this system.  The undoped sample exhibits negative curvature in Fig. 6, and as reported earlier it does not appear to fit the Korringa relation.~\cite{Kotegawa,Me}  K$_{x}$Fe$_{2-y}$Se$_{1.2}$S$_{0.8}$ also exhibits positive curvature, which suggests that it does not fit in a simple Fermi liquid picture either.  The K$_{x}$Fe$_{2-y}$Se$_{0.4}$S$_{1.6}$ data may exhibit a somewhat closer to linear relationship, but this could be fictitious in view of the very weak temperature dependences of $^{77}K$ and 1/T$_{1}$T in the temperature range of our investigation.  In addition, extrapolation of the apparent linear fit to the intercept of the horizontal axis implies $K_{chem}$ is as large as $\sim$ 0.22\% for K$_{x}$Fe$_{2-y}$Se$_{0.4}$S$_{1.6}$.  That is, if the Fermi liquid picture applies to the non-superconducting K$_{x}$Fe$_{2-y}$Se$_{0.4}$S$_{1.6}$, $K_{chem}$ would have to double in magnitude compared with superconducting K$_{x}$Fe$_{2-y}$S$_{2}$. 


\section{Summary}
In this paper we have reported a systematic variation of the electronic properties of the recently discovered high $T_{c}$ superconductor, K$_{x}$Fe$_{2-y}$Se$_{2-z}$S$_{z}$ with z = 0, 0.8, and 1.6, where S substitution generates chemical pressure.  Using $^{77}$Se NMR, we have taken measurements of Knight shift, $^{77}K$, and nuclear spin-lattice relaxation rate, 1/T$_{1}$, in a temperature range of up to 250 K.  The NMR results observed for the superconducting z = 0.8 sample ($T_{c} \sim$ 26 K) are analogous to those observed for the z = 0 sample ($T_{c} \sim$ 33 K) and overdoped Ba(Fe$_{1-x}$Co$_{x}$)$_{2}$As$_{2}$.  Specifically, the pseudogap-like behaviour with growing $^{77}$K and 1/T$_{1}$T at higher temperatures is strongly suppressed by S substitution.  The pseudogap-like behaviour is nearly non-existent for the non-superconducting z = 1.6 sample.  None of the K$_{x}$Fe$_{2-y}$Se$_{2-z}$S$_{z}$ systems \textit{conclusively} satisfy the Korringa relation expected for canonical Fermi liquids, but the z = 1.6 composition may in fact be closer to a Fermi liquid.     

It has been suggested that in Fe-As systems, the pnictogen height, i.e. the distance between the pnictogen and the Fe plane, acts as a tuner for superconductivity, where a certain height optimizes $T_{c}$.~\cite{Kuroki}  In addition, in the FeSe$_{1-x}$S$_{x}$ system it was reported that $T_{c}$ increased with S substitution up to 20\%, then decreased with further substitution.~\cite{Mizuguchi3}  Given that S substitution only lowers $T_{c}$ in our samples,~\cite{Petrovic1,Petrovic4} along with our results showing suppression of spin susceptibility and spin excitations similar to overdoped Fe-based systems, this suggests that increasing S content is analogous to \textit{over-pressurizing} the system beyond the optimum Fe-Se height.     

\begin{acknowledgements}
Work at McMaster is supported by NSERC and CIFAR. Work at Brookhaven is supported
by the U.S. DOE under Contract No. DE-AC02-98CH10886 and in part by the Center for
Emergent Superconductivity, an Energy Frontier Research Center funded by the U.S. DOE,
Office for Basic Energy Science (H.C.Lei and C.P).

\end{acknowledgements}           

\end{document}